\documentclass[]{article}
\usepackage{amssymb}
\def\ba{\mathbf{a}}   %\def\ba{{\bf a}}
\def\bb{\mathbf{b}}   %\def\bb{{\bf b}}
\def\be{\mathbf{e}}   %\def\bd{{\bf e}}
\def\bm{\mathbf{m}}   %\def\bm{{\bf m}}
\def\bn{\mathbf{n}}   %\def\bn{{\bf n}}
\def\bx{\mathbf{x}}   %\def\bx{{\bf x}}
\def\by{\mathbf{y}}   %\def\by{{\bf y}}
\def\bJ{\mathbf{J}}   %\def\bJ{{\bf J}}
\def\bS{\mathbf{S}}   %\def\bS{{\bf S}}
\def\bK{\mathbf{K}}   %\def\bK{{\bf K}}
\def\bM{\mathbf{M}}
\def\bN{\mathbf{N}}   %\def\bN{{\bf N}}

\def\C{\mathbb{C}} %\def\C{I\!\!\!C} %\def\C{{C\kern-.647em I}}
\def\G{\mathbb{G}} % \def\G{I\!\!\!G}

\def\R{\mathbb{R}}  %\def\R{I\!\!R}

\def\no{\noindent}

\def\beq{\begin{equation}}
\def\eeq{\end{equation}}
\def\con{\overline}

\def\w{\wedge}

\usepackage{helvet}
\usepackage{courier}
\usepackage{graphicx}   
\usepackage{makeidx}
\usepackage{multicol} % used for the two-column index
\usepackage{mathptmx} %**
\def\bpm{\begin{pmatrix}}
\def\epm{\end{pmatrix}}
\makeindex            
\begin{document}
\title{Part II: Spacetime Algebra of Dirac Spinors}
\author{Garret Sobczyk
\\ Universidad de las Am\'ericas-Puebla
 \\ Departamento de F\'isico-Matem\'aticas
\\72820 Puebla, Pue., M\'exico
\\ http://www.garretstar.com}
\maketitle
\begin{abstract} In Part I: {\it Vector Analysis of Spinors},
 the author studied the geometry of two component spinors as points
  on the Riemann sphere in the geometric algebra $\G_3$ of three dimensional 
  Euclidean space. Here, these ideas are generalized to apply to
  four component Dirac spinors on the complex Riemann sphere in 
  the complexified geometric algebra $\G_3(\C)$ of spacetime, which includes
Lorentz transformations. The
  development of generalized Pauli matrices 
  eliminate the need for the traditional Dirac gamma matrices.
 We give the discrete probability distribution of measuring a spin $1/2$
  particle in an arbitrary spin state, assuming that it was prepared in a given state immediately prior to
the measurement, independent of the inertial system in which measurements are made. The Fierz identities
between the physical observables of a Dirac spinor are discussed.

\smallskip
\no {\em AMS Subject Classication:} 15A66, 81P16 
\smallskip

\no {\em Keywords:} bra-ket formalism, geometric algebra, spacetime algebra, Dirac equation, 
Dirac-Hestenes equation, Riemann sphere, complex Riemann sphere, spinor, spinor operator, Fierz identities.

\end{abstract}

\section*{0 \ \ \  Introduction} 

Since the birth of quantum mechanics a Century ago, scientists have been both puzzled and amazed about the
seemingly inescapable occurrence of the imaginary number $i=\sqrt{-1}$, first in the Pauli-Schr\"odinger equation
for spin $\frac{1}{2}$ particles in space, and later in the more profound Dirac equation of spacetime. Exactly what 
role complex numbers play in quantum mechanics is even today hotly debated. In a previous paper, ``Vector Analysis of 
Spinors", I show that the $i$ occurring in the Schr\"odinger-Pauli equation for the electron should be interpreted
as the unit pseudoscalar, or directed volume element, of the geometric algebra $\G_3$.  
This follows directly from the assumption that the famous Pauli matrices are nothing more than the components
of the orthonormal space vectors $\be_1, \be_2, \be_3 \in \R^3$ with respect to the {\it spectral basis} of the
geometric algebra $\G_3$, \cite{S2014}. Another basic assumption made is that the geometric algebra $\G_3$ of space 
is naturally identified as the even sub-algebra $\G_{1,3}^+$ of the spacetime algebra $\G_{1,3}$, also known as the
algebra of {\it Dirac matrices}, \cite{H66}, \cite{S81}.

This line of research began when I started looking at the foundations of quantum mechanics. In particular,
I wanted to understand in exactly what sense the {\it Dirac-Hestenes} equation for the electron is equivalent to the
standard Dirac equation. What I discovered was that the equations are equivalent only so long as the issues of parity and complex
conjugation are not taken into consideration, \cite{S1/2}.
In the present work, I show that $i=\sqrt{-1}$ in the Dirac equation must have a different interpretation, than the $i$
that occurs in simpler Schr\"odinger-Pauli theory. In order to turn both the Schr\"odinger-Pauli theory, and the relativistic Dirac
theory, into strictly equivalent geometric theories, we replace the study of 2 and 4-component spinors with corresponding
2 and 4-component geometric spinors, defined by the minimal left ideals in the appropriate geometric algebras. As
pointed out by the late Perrti Lounesto, \cite[p.327]{LP97}, ``Juvet 1930 and Sauter 1930 replaced column spinors by square matrices in which only
the first column was non-zero - thus spinor spaces became minimial left ideals in a matrix algebra". In order to gives the resulting
matrices a unique geometric interpretation, it is then only necessary to interpret these matrices as the components of geometric
numbers with respect to the {\it spectral basis} of the appropriate geometric algebra \cite[p.205]{SNF}.  

The important role played by an idempotent, and its interpretation as a point on the Riemann sphere in the case of Pauli spinors, and as
a point on the complex Riemann sphere in the case of Dirac spinors,  make up the heart of our new geometric theory.
Just as the spin state of an electron can be identified with a point on the Riemann sphere, and a corresponding unique
point in the plane by stereographic projection from the South Pole, we find that the spin state of a relativistic electron can
be identified by a point on the complex Riemann sphere, and its corresponding point in the complex 2-plane by a
{\it complex stereographic projection} from the South Pole. In developing this theory, we find that the study of geometric Dirac spinors
can be carried out by introducing a generalized set of $2\times 2$ {\it Pauli E-matrices} over a 4-dimensional commuative ring
with the basis $\{1, i , I, i I\}$, where $i=\sqrt{-1}$ and $I=\be_{123}$ is the unit pseudo-scalar of the geometric algebra $\G_3$.
The setting for the study of quantum mechanics thereby becomes the complex geometric algebra $\G_3(\C)$. In order to
study quantum mechanics in a real geometric algebra, eliminating the need for any artificial $i=\sqrt{-1}$, we would have to
consider at least one of higher dimensional geometric algebras $\G_{2,3}, \G_{4,1}, \G_{0,5}$ of the respective
pseudoeuclidean spaces $\R^{2,3}, \R^{4,1}, \R^{0,5}$, \cite[p.217]{LP97}, \cite[p.326]{H85}.

\section{Geometric algebra of spacetime}

 The geometric algebra $\G_3$ of an orthonormal rest frame $\{\be_1,\be_2,\be_3\}$ in $\R^3$
 can be factored into an orthonormal frame $\{\gamma_0, \gamma_1,\gamma_2, \gamma_3\}$ 
 in the geometric algebra $\G_{1,3}$ of the pseudo-Euclidean space $\R^{1,3}$ of
 {\it Minkowski spacetime}, by writing 
  \beq  \be_k := \gamma_k \gamma_0=-\gamma_0 \gamma_k \quad {\rm  for} \quad k=1,2,3. \label{restframegamma} \eeq
 In doing so, the geometric algebra $\G_3$ is identified with the elements of the even sub-algebra $\G_{1,3}^+ \subset \G_{1,3}$.
 A consequence of this identification is that space vectors $\bx = x_1 \be_1+x_2 \be_2 + x_3 \be_3 \in \G_3^1$ become
{\it spacetime bivectors} in $\G_{1,3}^2\subset \G_{1,3}^+$. In summary, the geometric algebra $\G_{1,3}$, also known as {\it spacetime
 algebra} \cite{H66}, has $2^4=16$ basis elements generated by geometric multiplication of the $\gamma_\mu$ for $\mu=0,1,2,3$. Thus,
 \[\G_{1,3} := gen\{\gamma_0,\gamma_1,\gamma_2,\gamma_3\}  \]
 obeying the rules
 \[ \gamma_0^2=1, \ \gamma_k^2 = -1, \ \gamma_{\mu \nu}:=\gamma_\mu \gamma_\nu = -\gamma_\nu \gamma_\mu =\gamma_{\nu \mu} \]
 for $\mu \ne \nu$, $\mu, \nu = 0,1,2,3$, and $k=1,2,3$. Note also that the {\it pseudo-scalar} 
 \[\gamma_{0123}:=\gamma_{10}\gamma_{20}\gamma_{30}=\be_1 \be_2 \be_3=\be_{123}=:I \]
 of $\G_{1,3}$ is the same as the pseudo-scalar of the rest frame $\{\be_1, \be_2, \be_3\}$ of $\G_3$, and it
  anti-commutes
 with each of the spacetime vectors $\gamma_\mu$ for $\mu = 0,1,2,3$. 
 
    In the above, we have carefully distinguished the rest frame $\{\be_1, \be_2,\be_3\}$ 
    of the geometric algebra $\G_3  :=\G_{1,3}^+$.
 Any other rest frame $\{\be_1^\prime,\be_2^\prime,\be_3^\prime\}$
 can be obtained by an ordinary space rotation of the rest frame $\{\be_1, \be_2,\be_3\}$ followed
 by a {\it Lorentz boost}. In the spacetime algebra $\G_{1,3}$, this is equivalent to defining a new frame of
 spacetime vectors $\{\gamma_\mu^{\,\prime}| \ 0 \le \mu \le 3\} \subset \G_{1,3}$, and
 the corresponding rest frame $\{\be^\prime_k=\gamma_k^{\,\prime} \gamma_0^{\,\prime}| \ k=1,2,3 \}$ of a Euclidean space
 $\R^{3^\prime}$ moving with respect to the Euclidean space $\R^3$ defined by the rest frame $\{\be_1,\be_2,\be_3\}$.
 Of course, the primed rest-frame $\{\be_k^\prime \}$, itself, generates a corresponding geometric algebra $\G_3^\prime := \G_{1,3}^+$.
 A much more detailed treatment of $\G_3$ is given in \cite[Chp.3]{SNF}, and in \cite{S08} I explore the close
relationship that exists between geometric algebras and their matrix counterparts. 
  The way we introduced the geometric algebras $\G_3$ and $\G_{1,3}$ may appear novel, but they perfectly reflect all the
  common relativistic concepts \cite[Chp.11]{SNF}. 
    
 The well-known {\it Dirac matrices} can be obtained as a real sub-algebra of the $4\times 4$ matrix algebra $Mat_\C(4)$ over
 the complex numbers where $i=\sqrt{-1}$.
   We first define the idempotent 
 \beq u_{++}:=\frac{1}{4}(1+\gamma_0)(1+i\gamma_{12})=\frac{1}{4}(1+i\gamma_{12})(1+\gamma_0) ,\label{uplusplus} \eeq
 where the unit imaginary $i=\sqrt{-1}$ is assumed to commute with all elements of $\G_{1,3}$. Whereas it would be
 nice to identify this unit imaginary $i$ with the pseudo-scalar element $\gamma_{0123}=\be_{123}$ as we did in $ \G_3$,
  this is no longer possible since
 $\gamma_{0123}$ anti-commutes with the spacetime vectors $\gamma_\mu$ as previously mentioned. 
 %However,
 %the pseudo-scalar element $i=\gamma_{0^\prime 0123}$ of $\G_{2,3}$ {\it does} have the required property since it
 %is in the center of $\G_{2,3}$. 
 
 Noting that 
 \[ \gamma_{12}=\gamma_1 \gamma_0\gamma_0 \gamma_2 =\be_2 \be_1 = \be_{21}, \] 
 and similarly $\gamma_{31}= \be_{13}$, it follows that  
 \beq  \be_{13} u_{++} = u_{+-}\be_{13}, \ \  \be_{3} u_{++} = u_{-+}\be_{3}, \ \  \be_{1} u_{++} = u_{--}\be_{1}, \label{propidempotent}
 \eeq
 where
 \[u_{+-}:=\frac{1}{4}(1+\gamma_0)(1-i\gamma_{12}), \  u_{-+}:=\frac{1}{4}(1-\gamma_0)(1+i\gamma_{12}), \ 
  u_{--}:=\frac{1}{4}(1-\gamma_0)(1-i\gamma_{12}). \]
  The idempotents $u_{++},\, u_{+-}, \, u_{-+}, \, u_{--}$ are {\it mutually annihilating} in the sense that
  the product of any two of them is zero, and {\it partition unity} 
  \beq  u_{++}+u_{+-}+u_{-+}+u_{--} =1.\label{muanidepo} \eeq
  
  By the {\it spectral basis} of the Dirac algebra $\G_{1,3}$, we mean the elements of the matrix
\beq \pmatrix{1 \cr \be_{13} \cr \be_3 \cr \be_1} u_{++}  \pmatrix{1 & -\be_{13} & \be_3 & \be_1}
= \pmatrix{u_{++} & -\be_{13}u_{+-} & \be_3 u_{-+} & \be_1 u_{--}
 \cr \be_{13} u_{++} &u_{+-} & \be_1 u_{-+} & -\be_3 u_{--} \cr
 \be_3 u_{++} & \be_{1}u_{+-} & u_{-+} & -\be_{13} u_{--} \cr 
  \be_1 u_{++} & - \be_{3}u_{+-} & \be_{13} u_{-+} &  u_{--} }. \label{specbasisD} \eeq
  Any geometric number $g \in \G_{1,3}$ can be written in the form
  \beq  g = \pmatrix{1 & \be_{13} & \be_3 & \be_1} u_{++} [g] \pmatrix{1 \cr -\be_{13} \cr \be_3 \cr \be_1} \label{anyg} \eeq
  where $[g]$ is the {\it complex Dirac matrix} corresponding to the geometric number $g$. In particular,
  \beq [\gamma_0] = \pmatrix{1 & 0 & 0 & 0 \cr 0 & 1 & 0 & 0 \cr 0 & 0 & -1 & 0 \cr 0 & 0 & 0 & -1}, 
   [\gamma_1] =\pmatrix{0 & 0 & 0 & -1 \cr 0 & 0 & -1 & 0 \cr 0 & 1 & 0 & 0 \cr 1 & 0 & 0 & 0},
   \label{gammamat} \eeq
 and
  \[[\gamma_2] =\pmatrix{0 & 0 & 0 & i \cr 0 & 0 & -i & 0 \cr 0 & -i & 0 & 0 \cr i & 0 & 0 & 0},
   [\gamma_3] =\pmatrix{0 & 0 & -1 & 0 \cr 0 & 0 & 0 & 1 \cr 1 & 0 & 0 & 0 \cr 0 & -1 & 0 & 0}.\]

    It is interesting to see what the representation is of the basis vectors of $\G_3$. We find that for
 $k=1,2,3$,
 \[ [\be_k]_4=[\gamma_k][\gamma_0]=\pmatrix{[0]_2 & [\be_k]_2 \cr [\be_k]_2 & [0]_2 } \quad {\rm and} 
 \quad [\be_{123}]_4 = i \pmatrix{[0]_2 & [1]_2 \cr [1]_2 & [0]_2 },    \]
  where the outer subscripts denote the order of the matrices and, in particular, $[0]_2$, $[1]_2$ are 
  the $2\times2$ zero and unit matrices, respectively.
 The last relationship shows that the $I:=\be_{123}$ occurring in the Pauli matrix representation,
 which represents the oriented unit of volume, is different than the $i=\sqrt{-1}$ which occurs in
 the complex matrix representation (\ref{gammamat}) of the of Dirac algebra. In particular,
 $[\be_2]_2 := \pmatrix{0 & -i \cr i & 0}$, which is {\bf not} the Pauli matrix for $\be_2\in \G_3$ since $i \ne I$.
 We will have more to say about this important matter later.
 
 A {\it Dirac spinor} is a $4$-component column matrix $[\varphi]_4$,
 \beq [\varphi]_4:=\pmatrix{\varphi_1 \cr \varphi_2 \cr \varphi_3 \cr \varphi_4} \quad {\rm for}\quad
    \varphi_k =x_k+iy_k \in \C . \label{diracspinorop} \eeq
 Just as in \cite{S1/2}, from the Dirac spinor $[\varphi]_4$, using (\ref{anyg}),
  we construct its equivalent $S\in \G_{1,3}(\C)$ as an element of the {\it minimal left ideal} generated by $u_{++}$,
 \beq  [\varphi]_4=\pmatrix{\varphi_1 \cr \varphi_2 \cr \varphi_3 \cr \varphi_4} \leftrightarrow 
     \pmatrix{\varphi_1 & 0 & 0 &0 \cr \varphi_2  & 0 & 0 &0 \cr \varphi_3  & 0 & 0 &0 \cr \varphi_4 & 0 & 0 &0 }
      \leftrightarrow S:=\big(\varphi_1 +\varphi_2 \be_{13}+\varphi_3 \be_3+\varphi_4 \be_1\big)u_{++} .
                         \label{diracspinor}         \eeq
     Because of its close relationship to a Dirac spinor, we shall refer to $S$ as a {\it geometric Dirac spinor} or a
     {\it Dirac g-spinor}.                    
     
 Noting that 
 \[u_{++} \gamma_{21}=\frac14 (1 + \gamma_0) (\gamma_{21} + i \gamma_{12} \gamma_{21}) 
 = \frac14 (1 + \gamma_0) (i - \gamma_{12}) =iu_{++}=\gamma_{21}u_{++},\]
  it follows that $ \varphi_k u_{++} =(x_k   +   \gamma_{21} y_k) u_{++} = u_{++} 
  \big(x_k + \gamma_{21} y_k\big)$ and hence
 \beq S  =\big(\alpha_1 +\be_{13}\alpha_2 +\be_3\alpha_3 +\be_1\alpha_4 \big)u_{++} 
  =\big(\alpha_1 +\alpha_2^\dagger \be_{13} +\be_3\alpha_3 +\alpha_4^\dagger\be_1 \big)u_{++} , \label{Salpha} \eeq
 where each of the elements $\alpha_k$ in $S$ is defined by $\alpha_k=\varphi_k|_{i\to \gamma_{21}}$, 
 and $\alpha_k^\dagger:=\varphi_k|_{i\to -\gamma_{21}}$. 
 
 Expanding out the terms in (\ref{Salpha}),
 \beq S =\Big(( x_1+x_4 \be_1 + y_4 \be_2+x_3 \be_3)+I(y_3+y_2 \be_1-x_2 \be_2 + y_1 \be_3)\Big)u_{++} \label{Salphaexpand} .\eeq
 This suggest the substitution
 \beq \varphi_1 \to x_0+i y_3,\ \ \varphi_2 \to -y_2+i y_1,\ \ \varphi_3 \to x_3+i y_0, \ \ \varphi_4 \to x_1+i x_2 ,
             \label{substitutionxy} \eeq
   in which the geometric Dirac spinor $S$ takes the more perspicuous forms  
 \beq S =\big( X +I\,Y\big)u_{++}=\big( X +I\,Y\big)\gamma_0 u_{++}=(x+I\,y)u_{++} , \label{Salphaexpandnew} \eeq  
 for $X:=x_0+\bx,\ Y:=y_0+\by \in \G_3$ where $\bx=x_1 \be_1 + x_2 \be_2+x_3 \be_3,\  \by=y_1 \be_1+y_2 \be_2 + y_3 \be_3$,
 and $x:=\sum_{\mu=0}^3 x_\mu \gamma_\mu, y:=\sum_{\mu=0}^3 y_\mu \gamma_\mu\in \G_{1,3}$.
         
  We now calculate
  \[ \con S=\big(\alpha_1 +\be_{13}\alpha_2 +\be_3\alpha_3 +\be_1\alpha_4 \big)u_{+-} ,  \]
  where $\con S$ is the {\it complex conjugate} of $S$, defined by $i\to -i$,   
    \[ S^{\#}=\big(\alpha_1 +\be_{13}\alpha_2 +\be_3\alpha_3 +\be_1\alpha_4 \big)u_{-+} , \]
  where $S^{\#}$ is the parity transformation defined by $\gamma_\mu \to -\gamma_\mu$, and
   \[ S^\star:=(\con S)^{\#}=\big(\alpha_1 +\be_{13}\alpha_2 +\be_3\alpha_3 +\be_1\alpha_4 \big)u_{--} . \]

   Using (\ref{muanidepo}), we then define the {\it even spinor operator}
\[ \psi :=S+\con S+ S^{\#} + S^\star=\big(\alpha_1 +\be_{13}\alpha_2 +\be_3\alpha_3 +\be_1\alpha_4 \big)\big(u_{++}+u_{+-}
+u_{-+}+u_{--}\big) \]   
\beq = \alpha_1 +\be_{13}\alpha_2 +\be_3\alpha_3 + 
  \be_1\alpha_4 =X+ I\,Y \in \G_{1,3}^+, \label{relatepsi} \eeq
   and the {\it odd spinor operator}
 \[ \Phi :=S+\con S- S^{\#} - S^\star=\big(\alpha_1 +\be_{13}\alpha_2 +\be_3\alpha_3 +\be_1\alpha_4 \big)\big(u_{++}+u_{+-}
-u_{-+}-u_{--}\big) \] 
 \beq =\big( \alpha_1 +\be_{13}\alpha_2 +\be_3\alpha_3 + 
  \be_1\alpha_4 \big)\gamma_0=x+ I\,y \in \G_{1,3}^-. \label{oddrelatepsi} \eeq 
  In addition to the two real even and odd spinor operators in $\G_{1,3}$,
  we have two {\it complex spinor operators} in $\G_{1,3}(\C)$, given by
  \[ Z_+ :=S-\con S+ S^{\#} - S^\star=\big(\alpha_1 +\be_{13}\alpha_2 +\be_3\alpha_3 +\be_1\alpha_4 \big)\big(u_{++}-u_{+-}
+u_{-+}-u_{--}\big) \]
     \beq    =  \big(\alpha_1 +\be_{13}\alpha_2 +\be_3\alpha_3 +\be_1\alpha_4 \big) E_3=(X+ I\,Y)E_3 \in \G_{1,3}^+(\C) ,  
            \label{complexspinor1} \eeq
  where $E_3:=-iI\be_3$, and  
   \[ Z_- :=S-\con S - S^{\#} + S^\star=\big(\alpha_1 +\be_{13}\alpha_2 +\be_3\alpha_3 +\be_1\alpha_4 \big)\big(u_{++}-u_{+-}
-u_{-+}+u_{--}\big) \]
     \beq    =  \big(\alpha_1 +\be_{13}\alpha_2 +\be_3\alpha_3 +\be_1\alpha_4 \big)\gamma_0E_3
     =(x+I\,y)E_3\in \G_{1,3}^-(\C) .  
            \label{complexspinor2} \eeq

    Using (\ref{specbasisD}), the matrix $[\psi]$ of the even spinor operator $\psi$ is found to be
  \beq [\psi]= \pmatrix{\varphi_1 & -\con\varphi_2 & \varphi_3 &\con\varphi_4 \cr                
   \varphi_2 & \con\varphi_1 & \varphi_4 & -\con\varphi_3 \cr 
   \varphi_3 & \con\varphi_4 & \varphi_1 & -\con\varphi_2 \cr 
   \varphi_4 & -\con\varphi_3 & \varphi_2 & \con\varphi_1 }  ,          \label{dspinop} \eeq
 \cite{S1/2}, \cite[p.143]{LP97}, and using (\ref{gammamat}) and (\ref{dspinop}), 
the matrix $[\Phi]$ of the odd spinor operator $\Phi$ is found to be
 \beq [\Phi]= \pmatrix{\varphi_1 & -\con\varphi_2 & -\varphi_3 &-\con\varphi_4 \cr                
   \varphi_2 & \con\varphi_1 & -\varphi_4 & +\con\varphi_3 \cr 
   \varphi_3 & \con\varphi_4 & -\varphi_1 & +\con\varphi_2 \cr 
   \varphi_4 & -\con\varphi_3 & -\varphi_2 & -\con\varphi_1 }  .          \label{odddspinop} \eeq  

    Unlike the Dirac spinor $[\varphi]_4$, the even spinor operator $[\psi]$ is invertible iff $\det[\psi]\ne 0$.
 We find that
 \beq  \det[\psi] = r^2 +4 a^2 \ge 0 ,  \label{detpsi} \eeq
 where 
 \[ r= |\varphi_1|^2+|\varphi_2|^2 - |\varphi_3|^2-|\varphi_4|^2 \quad {\rm and} \quad 
   a=im\, \big(\con \varphi_1 \varphi_3 + \con \varphi_2 \varphi_4\big). \]  
 Whereas the even spinor operator $[\psi]$ obviously contains the same information as the Dirac spinor
 $[\varphi]_4$, it acquires in (\ref{relatepsi}) the geometric interpretation of an even multivector in $\G_{1,3}^+$. 
 %Using (\ref{anyg}),
 % the  geometric number $\psi \in \G_{1,3}^+$ corresponding to the matrix spinor operator $[\psi]$ is
 %\[ \psi = \pmatrix{1 & \be_{13} & \be_3 & \be_1} u_{++} [\psi]_\alpha \pmatrix{1 \cr -\be_{13} \cr \be_3 \cr \be_1} \]
 % \beq = \alpha_1 +\be_{13}\alpha_2 +\be_3\alpha_3 +\be_1\alpha_4 =  \alpha_1 
 % +\con\alpha_2\be_{13} +\alpha_3\be_3 +\con\alpha_4\be_1
 %  \in \G_{1,3}^+$\Phi$. \label{gspinoreven} \eeq 
 With the substitution (\ref{substitutionxy}), the expansion of the determinant (\ref{detpsi}) takes the
 interesting form
 \[ \det[\psi] = \big(( x_0^2-x_1^2 - x_2^2- x_3^2)-(y_0^2-y_1^2 -y_2^2 - y_3^2)\big)^2+4(x_0y_0-x_1y_1-x_2y_2-x_3y_3)^2 \]  
  \beq =(x^2-y^2)^2+4(x\cdot y)^2 \quad {\rm for } \quad x,y \in \G_{1,3}^1. \label{detpsixy} \eeq
 Since the matrix $[\Phi]$ of the odd spinor operator is the same as the matrix $[\psi]$ of the even spinor operator,
 except for the change of sign in the last two columns, $\det[\Phi]=\det[\psi]$. Indeed, similar arguments apply
 to the matrices $[Z_+]$ and $[Z_-]$, and so $\det[\psi]=\det[Z_+]=\det[Z_-]$. 
   
As an even geometric number in $\G_{1,3}^+ $, $\psi$ generates Lorentz boosts in
addition to ordinary rotations in the Minkowski space $\R^{1,3}$. 
Whereas we started by formally introducing the complex
number $i=\sqrt{-1}$, in order to represent the Dirac gamma matrices, we have ended up with the real spinor operators
$\psi, \Phi \in \G_{1,3}^+$, in which the role of $i=\sqrt{-1}$ is taken over by the $\gamma_{21}=\be_{12}\in \G_3$.
This is the key idea in the Hestenes representation 
of the {\it Dirac equation} \cite{H10}. However, in the process, the crucial role played by the mutually annihilating idempotents $u_{\pm \pm}$
has been obscured, and the role played by $i=\sqrt{-1}$ in the definition of the Dirac matrices has been buried.
Idempotents are slippery objects which can change the identities of everything they touch. As such, they should always be
treated gingerly with care. Idempotents naturally arise in the study of number systems that have zero divisors, \cite{S1}.
    
 \section{Geometric Dirac Spinors to Geometric E-Spinors}

Recall from equation (\ref{relatepsi}) that the real, even, Dirac spinor operator is
\[ \psi = \alpha_1 + \be_{13}\alpha_2+ \be_3\alpha_3+\be_1\alpha_4 =X+I\,Y \in \G_{1,3}^+, \]
for the geometric Dirac spinor,
 \[ S=\psi u_{++}=\Big(\varphi_1 +\varphi_2 \be_{13}+\varphi_3 \be_3+\varphi_4 \be_1\Big)u_{++}=
\Big( \alpha_1 + \be_{13}\alpha_2+ \be_3\alpha_3+\be_1\alpha_4\Big)u_{++}  \]
as an element in the minimal left ideal $\{\G_{1,3}^+u_{++}\}$.
 
 Defining $J:=-iI$, we can express  
 \[ u_{++}=\gamma_0^+ E_3^+, \ \ {\rm where} \ \ \gamma_0^\pm=\frac{1}{2}(1\pm \gamma_0), \ \ 
   E_3^\pm = \frac{1}{2}(1 \pm J\be_3) .\]  
    Noting that $J\be_3 u_{++}=u_{++}$, 
   \[ S=\Big(\varphi_1 +\varphi_2 \be_{13}+\varphi_3 \be_3+\varphi_4 \be_1\Big)u_{++} 
    = \Big(\varphi_1 +\varphi_2 \be_{1}J+\varphi_3 J+\varphi_4 \be_1\Big)u_{++} \]
\beq = \Big((\varphi_1 +\varphi_3J)+(\varphi_4 +\varphi_2 J) \be_1\Big)u_{++}
 =\Omega u_{++},\label{Mdef} \eeq
 where $\Omega:=\Omega_0+\Omega_1 \be_1$ for $\Omega_0:=(\varphi_1 +\varphi_3J)$ and $\Omega_1:=(\varphi_4 +\varphi_2J )$.  
Alternatively, $\Omega_0$ and $\Omega_1$ can be defined in the highly useful, but equivalent, way
\beq \Omega_0 = z_1+J z_3, \quad {\rm and} \quad \Omega_1= z_4+ z_2 J, \label{equivOmega} \eeq
for $z_1:=x_1+y_3 I,\ z_3:= x_3+y_1 I,\ z_4:=x_4+y_2 I,\ z_2:= x_2+y_4I$ are all in $\G_{1,3}^{0+4}$. With the substitution
variables (\ref{substitutionxy}), the $z_k $'s become 
\beq z_1=x_0+y_0 I,\ \ z_2= -y_2+x_2 I,\ \ z_3= x_3+y_3 I,\ \ z_4=x_1+y_1 I . \label{equivOmegasub} \eeq

Once again, we calculate
  \[ \con S=\Big((\con\varphi_1 -\con\varphi_3J)  +(\con\varphi_4 -\con\varphi_2J )\be_{1} \Big)u_{+-}=
     \con \Omega u_{+-},  \]
  where $\con S$ is the {\it complex conjugate} of $S$ defined by $i\to -i$,   
    \[ S^{\#}=\Big((\varphi_1 +\varphi_3J)  +(\varphi_4 +\varphi_2J )\be_{1}\Big)u_{-+}=\Omega u_{-+}, \]
  where $S^{\#}$ is the parity transformation defined by $\gamma_\mu \to -\gamma_\mu$, and
   \[ S^\star:=(\con S)^{\#}=\Big((\con\varphi_1 -\con\varphi_3J)  +(\con\varphi_4 - \con\varphi_2J )\be_{1} \Big)u_{--}=
       \con \Omega u_{--}. \]
The even spinor operator (\ref{relatepsi}), satisfies
\beq \psi =S+\con S+ S^{\#} + S^\star=\Omega E_3^+ +\con \Omega E_3^-=\frac{1}{2}(\Omega+\con \Omega)
+\frac{1}{2}(\Omega-\con \Omega)E_3, \label{relatepsinew} \eeq
 where as before $E_3:=J\be_3=E_3^+-E_3^-$.   
  Equations (\ref{relatepsi}) and (\ref{relatepsinew}) express the even spinor operator $\psi$
   in two very different, but
  equivalent ways, the first as an element of $\G_{1,3}^+$, and the second by multiplication in the complex geometric algebra
  $\G_{1,3}^+(\C)$. The second approach is closely related to the {\it twistor theory} of Roger Penrose \cite[p.974]{Pen04}.
  
%   We now explore the {\it twistor} 
%\beq S=\psi u_{++}=Ru_{++}  = \Big((\varphi_1 +\varphi_3J)+(\varphi_4 +\varphi_2 J) \be_1\Big)u_{++}
%=\Big(\Omega_0+\Omega_1\be_1\Big)u_{++}, \label{spintwistor} \eeq
%where $\Omega_0:=(\varphi_1 +\varphi_3J)$ and  $\Omega_1:=(\varphi_4 +\varphi_2J)$. 
 It follows from  (\ref{Mdef}) and (\ref{relatepsinew}), that any even spinor operator $\psi\in \G_{1,3}^+$ can be written in the matrix form
 \beq \psi=\pmatrix{1 & \be_1} E_3^+ \pmatrix{\Omega_0 & \con\Omega_1 \cr \Omega_1 & \con \Omega_0}\pmatrix{1 \cr \be_1} \ \ \iff  \ \
        \psi^\dagger =\pmatrix{1 & \be_1} E_3^+ \pmatrix{\con \Omega_0^\dagger & \con\Omega_1^\dagger
         \cr \Omega_1^\dagger & \Omega_0^\dagger }\pmatrix{1 \cr \be_1}
                         \label{twistorrep} \eeq
 where the matrix $[\psi]_\Omega
 := \pmatrix{\Omega_0 & \con\Omega_1 \cr \Omega_1 & \con \Omega_0}$, and 
 the matrix $[\psi^\dagger]_\Omega
 := \pmatrix{\con \Omega_0^\dagger & \con\Omega_1^\dagger
         \cr \Omega_1^\dagger & \Omega_0^\dagger }$.
  Analogous to the Pauli matrices, we have
  \beq [1]_\Omega =\pmatrix{1 & 0 \cr 0& 1}, \ \ [\be_1]_\Omega=\pmatrix{0 & 1 \cr 1& 0}, \ \ 
  [\be_2]_\Omega=\pmatrix{ 0 & -i \cr i& 0}, \ \ 
  [\be_3]_\Omega=\pmatrix{J & 0 \cr 0& -J} ,\label{analogpauli} \eeq
  which can be obtained from the Pauli matrix representation by multiplying the Pauli matrices for $\be_2$ and $\be_3$ by 
  $J=-i I$. Alternatively, the Pauli matrices can be obtained from the above Dirac-like 
  representation, simply by replacing $i$ by $I$, in which case $J\to 1$.  Indeed, if we let $i\to I$ in (\ref{twistorrep}), and using the
  substitution variables $z_k$ defined in (\ref{equivOmegasub} ), we get 
  \beq  \psi=\pmatrix{1 & \be_1} u_+ \pmatrix{z_1+z_3  & z_4-z_2 \cr z_4 +z_2 & z_1-z_3}\pmatrix{1 \cr \be_1},
                         \label{paulirep} \eeq
   where $u_+=\frac{1}{2}(1+\be_3)$, which is exactly the Pauli algebra representation of the geometric number $\psi \in \G_3$.  
   Recently, I was surprised to discover that I am not the first to consider such a representation of the Pauli matrices \cite{BK12}.                    
  
 Calculating the determinant of the matrix $[\Omega]:=[\psi]_\Omega$ 
 of $\psi$,
   \beq \det [\Omega] = \det \pmatrix{\Omega_0 & \con\Omega_1 \cr \Omega_1 & \con \Omega_0} 
   =|\varphi_1|^2  + |\varphi_2|^2 -|\varphi_3|^2 -|\varphi_4|^2 +
   2 im(\con\varphi_1\varphi_3+\con\varphi_2\varphi_4)I ,  \label{detomega} \eeq
 or in the more elegant alternative form in the spacetime algebra $\G_{1,3}$,
   \beq \det [\Omega] = x^2 - y^2 + 2 I (x\cdot y)=(x+I\,y)(x-I\,y)=(X+I\,Y)(\widetilde X+I\, \widetilde Y),
    \label{detomegaxy} \eeq
   where $x=X\gamma_0$ and $y=Y\gamma_0$, and $\widetilde A$ denotes the operation of reverse of the element $A\in \G_{1,3}$.
 Note that the determinant $\det[ \psi]$, found in (\ref{detpsi}), is related to (\ref{detomega}) or (\ref{detomegaxy}), by
  $\det[ \psi]=|\det[ \psi]_\Omega|^2$ for the matrix of the even spinor operator $\psi$
 defined in (\ref{dspinop}). The spinor operator $\psi$ will have an inverse only when $\det[\psi]\ne 0$. 
 %In terms of the substituted variables (\ref{substitutionxy}), the determinant (\ref{detomega}) takes the form
 %\beq \det [\psi]_\Omega 
 %  =( x_0^2-x_1^2 - x_2^2- x_3^2)-(y_0^2-y_1^2 -y_2^2 - y_3^2) +
 %  2 I (x_0y_0-x_1y_1-x_2y_2-x_3y_3) .  \label{detomegaxy} \eeq
  
 We have already noted that the geometric algebra $\G_3$ can be algebraically identified with the
 even sub-algebra $\G_{1,3}^+ \subset \G_{1,3}$. Each timelike Dirac vector $\gamma_0$ 
 determines a different rest frame (\ref{restframegamma}) of spacelike bivectors $\{\be_1,\be_2,\be_3\}$. 
 For $\psi\in \G_{1,3}^+$, the matrix $[\psi]_\Omega$ is {\it Hermitian} with respect to the rest-frame  $\{\be_1,\be_2,\be_3\}$ of
 $\gamma_0$ if 
 \[ \con{[\psi^\dagger]}_\Omega^T = [\psi]_\Omega ,\]
  or equivalently, $\psi^\dagger = \psi$, where $\dagger$ is the conjugation of reverse in the Pauli algebra
  $\G_3 \widetilde = \G_{1,3}^+$. Using the more transparent variables (\ref{substitutionxy}), 
 for a Hermitian $[\psi]_\Omega$,
  \[  [\psi]_\Omega=\pmatrix{\Omega_0 & \con\Omega_1 \cr \Omega_1 & \con \Omega_0} 
   =\pmatrix{x_0+Jx_3  & x_1-i x_2 \cr x_1+ix_2& x_0-Jx_3 } =\con{[\psi^\dagger]}_\Omega^T\]
  with $\det[\psi]_\Omega=x_0^2-x_1^2-x_2^2-x_3^2$, and for anti-Hermitian $[\psi]_\Omega$, 
  \[  [\psi]_\Omega=\pmatrix{\Omega_0 & \con\Omega_1 \cr \Omega_1 & \con \Omega_0}
=I\pmatrix{y_0+Jy_3 & y_1-iy_2 \cr y_1+iy_2&y_0-Jy_3} =- \con {[\psi^\dagger]}_\Omega^T\]
   with $\det[\psi]_\Omega=-y_0^2+y_1^2+y_2^2+y_3^2$.
    
    Thus, a {\it Hermitian Dirac g-spinor}, corresponding to $X=x_0 + \bx\in \G^3$, has the form 
  \[ S := (\Omega_0+\Omega_1 \be_1)u_{++}=\Big((x_0+x_3J)+(x_1+ix_2)\be_1 \Big) u_{++}=X\, u_{++}, \]
  for $\Omega_0=x_0+x_3J$ and $\Omega_1=x_1+ix_2$,
  and an anti-Hermitian Dirac g-spinor, corresponding to $I\, Y=I\,(y_0+ \by) \in \G_3$, has the form
  \[ Q:= (\Omega_0+\Omega_1 \be_1)u_{++} =  I\Big((y_0+y_3 J)+(y_1+i y_2)\be_1 \Big) u_{++}=I\,Y \,u_{++}, \] 
   for $\Omega_0= I(y_0+y_3 J)$ and $\Omega_1=I(y_1+iy_2)$.
   
  In the bra-ket notation, a Dirac g-spinor (\ref{diracspinor}), and its conjugate Dirac g-spinor, in the spacetime algebra $\G_{1,3}^+(\C)$, takes
  the form
   \beq  |\Omega\rangle := 2S=2(\Omega_0 + \Omega_1 \be_1 )u_{++}=2(x+Iy)u_{++} , \label{twistspin} \eeq
 and
   \beq  \langle\Omega | := \widetilde{\con{| \Omega \rangle}}=
     2u_{++}(\con\Omega_0 - \con\Omega_1 \be_1 )=2u_{++}(x-Iy), \label{contwistspin} \eeq  
 respectively. Taking the product of the conjugate of a g-spinor $|\Phi \rangle=2(r+Is)u_{++}$, with a g-spinor $|\Omega \rangle$, gives
  \[  \langle\Phi | |\Omega \rangle =4u_{++}\big( \con \Phi_0 - \be_1\con \Phi_1\big)\big(\Omega_0 + \be_1 \Omega_1 \big)u_{++}
    =4 u_{++}\big( \con \phi_1 \varphi_1+ \con \phi_2 \varphi_2  -\con \phi_3 \varphi_3 - \con \phi_4 \varphi_4\big)       \]
    \beq = 4u_{++}\Big(r\cdot x- s \cdot y +\gamma_{21}\big(\gamma_{12}\cdot(r\w x-s\w y)+\gamma_{30}\cdot(r\w y -s\w x)\big)\Big).
        \label{sesquipre}  \eeq

 The complex {\it sesquilinear inner product} in $\G_3^{0+3}$ of the two g-spinors $|\Phi \rangle$ and $|\Omega \rangle $ is then defined by
  \beq  \langle\Phi  |\Omega \rangle := \Big\langle  \langle\Phi  ||\Omega \rangle \Big\rangle_\C = 
   \Big\langle \con \phi_1 \varphi_1+ \con \phi_2 \varphi_2  -\con \phi_3 \varphi_3 - \con \phi_4 \varphi_4 \Big\rangle_\C 
   = a+ib \in \C.  
   \label{sesquip}  \eeq 
  Defining $|\Omega\rangle= 2(\Omega_0 + \Omega_1 \be_1 )u_{++} $ and $|\Phi \rangle=2(\Phi_0+\Phi_1\be_1)u_{++}$
  for the substituted variables (\ref{substitutionxy}), with $|\Phi \rangle$ being defined in the coresponding variable $r$ and $s$,
  the inner product (\ref{sesquip}) takes the form
  \[  \langle\Phi  |\Omega \rangle = (r_0x_0-r_1x_1-r_2x_2-r_3x_3)-(s_0 y_0-s_1 y_1-s_2 y_2-s_3y_3) \]
        \beq +    i(r_0y_3-r_3y_0+s_0x_3-s_3x_0+r_2x_1-r_1x_2+s_1y_2-s_2y_1).  
   \label{sesquipnew}  \eeq 
   Alternatively, writing $|\Omega\rangle =2(x+Iy)u_{++}$ and $\langle \Phi|=2u_{++}(r-Is)$, we have
   \[  \langle\Phi  |\Omega \rangle = (r\cdot x- s \cdot y )
   +i\big(\gamma_{12}\cdot(r\w x-s\w y)+\gamma_{30}\cdot(r\w y -s\w x)\big)\in    \C. \]
   
  Consider now the parity invarient part $S_{E}$ of the geometric Dirac spinor $S$,  
   \[ S_E:= S+S^{\#} = (\Omega_0+\Omega_1\be_1)E_3^+= \Omega E_3^+=(X+I\, Y)E_3^+.  \]
 Writing 
 \beq S_E=\Omega_0(1+\Omega_0^{-1}\Omega_1 \be_1)E_3^+ = \Omega_0 T_E^+ , \label{idempotenT} \eeq
 for $T_E^+:= (1+ \Lambda \be_1)E_3^+$ and  $\Lambda:= \Omega_0^{-1}\Omega_1$, we then find that
 $T_E^+$ is an idempotent. We now study idempotents $P \in \G_3(\C)$ of the form
 \beq P = \frac{1}{2}(1+ J  \bM+I \,\bN ), \label{complexidempotent} \eeq
 for $\bN, \bM\in \G_{1,3}^{2}$ and $( J  \bM+I \,\bN)^2=1$.  
 
 Defining the {\it symmetric product} $\bM\circ \bN:= \frac{1}{2}(\bM \bN+\bN \bM)$, the condition
   \[  ( J  \bM+I \,\bN)^2=1 \quad \leftrightarrow \quad \bM^2-\bN^2 = 1 \quad {\rm and} \quad \bM\circ \bN = 0, \]
implies that
  \[ P = \frac{1}{2}(1+J  \bM+I \,\bN) = J \,\bM \frac{1}{2}\Big( 1+ \frac{J}{\bM^2}\big(\bM+I\,\bM \bN\big)\Big)
     =J\,\bM \frac{1}{2}(1+ J\,\hat \bb) ,\]
 for $\hat \bb := \frac{1}{\bM^2}\big(\bM+I\,\bM \bN\big)  \in \G_{1,3}^2$. For the idempotent
 $T_E^+$, defined in (\ref{idempotenT}), we have 
 \beq T_E^+ =(1+ \Lambda \be_1)E_3^+ = \frac{1}{2}(1+J  \bM+I \,\bN)= J\,\bM E_3^+ = \bM \be_3 E_3^+,  \label{idempotentTcan}  \eeq
 since in this case $\hat \bb = \be_3$. 
  Because $T_E^+$ is an idempotent, it further follows that 
  \beq T_E^+ =J\,\bM E_3^+ = \bM^2(\hat \bM E_3^+ \hat \bM) E_3^+ =  \bM^2 \hat A_+ E_3^+, \label{idempotentTcan2} \eeq
 where $\hat A_+ :=(\hat \bM E_3^+ \hat \bM)$, and $\hat \bM := \frac{\bM}{\sqrt{\bM^2}}$. We also identify $J\hat \ba$
 in $\hat A_+$, by $\hat \ba = \hat \bM \be_3 \hat \bM\in \G_{1,3}^2$.  
 
\begin{figure}
\begin{center}
\no\includegraphics[scale=.50]{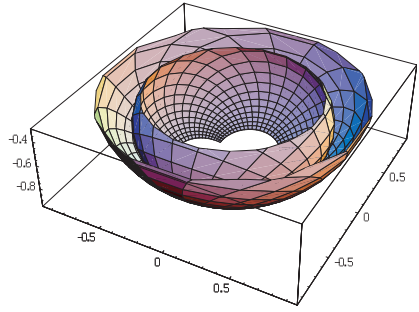}
\caption{The larger dish shows unit direction velocities on the Riemann sphere defined by $\hat \bM$.
 The inner dish shows the velocity vectors $\hat \bm_1\times \hat \bm_2 \tanh 2\phi$, measured from the
origin. The dishes coincide at the south pole, when $2\phi \to \infty$ and the velocity $v\to c$. }
\label{pick}
\end{center}
\end{figure}

 Now decompose the complex unit vector $\hat\bM=\bm_1+I \bm_2 \in \G_3^{1+2}$, by writing
 \beq \hat \bM =\hat \bm_1 \cosh \phi +I\hat \bm_2 \sinh \phi = e^{\phi \hat \bm_1\times \hat \bm_2}\hat \bm_1, 
  \label{boostsphereM} \eeq
 where $\cosh \phi =|\bm_1|$ and $\sinh \phi =|\bm_2|$,
  and note that since $\hat \bM^2 = 1$, 
  \[ \hat \bm_1\hat \bm_2 =-\hat \bm_2 \hat \bm_1= I (\hat \bm_1\times \hat \bm_2), \] 
  so $I=\be_{123}=\hat \bm_1 \hat \bm_2 (\hat \bm_1 \times \hat \bm_2)$.
  It then follows that the expression for $\hat \ba$, given after equation (\ref{idempotentTcan2}), becomes
 \beq  \hat \ba = \hat \bM \be_3 \hat \bM =e^{\phi \hat \bm_1\times \hat \bm_2}\hat \bm_1 \be_3 \hat\bm_1 
 e^{-\phi \hat \bm_1\times \hat \bm_2} ,\label{boostsphere} \eeq
 which expresses that the North Pole $\be_3$ of the Riemann sphere is rotated $\pi$ radians in the plane of $I\hat\bm_1$ into the
 point $\hat \bm_1 \be_3 \hat \bm_1$, and then undergoes the {\it Lorentz Boost} defined by the unit vector
 $\hat \bm_1 \times \hat \bm_2$, with velocity $\tanh 2\phi = v/c$, to give the unit vector $\hat \ba$. We can think of the
 Riemann sphere, itself, as undergoing a Lorentz boost with velocity $\hat \bm_1\times \hat \bm_2 \tanh 2\phi$. 
  See Figure \ref{pick}. 

 We can now decompose the parity invariant part $S_E$ of the geometric Dirac Spinor $S$, into
 \beq  S_E = \Omega_0 T_E^+ =J \Omega_0 \bM E_3^+ =\Omega_0  \bM^2 \hat A_3^+  E_3^+ , \label{decomposeSE} \eeq
          where $\Omega_0 =\varphi_1+J\varphi_3$, and 
        \[ T_E^+ =(1+\Lambda\, \be_1)E_3^+=\frac{1}{2}\big(1+J  \bM+I \,\bN\big), \]  
for $\Lambda = \Omega_0^{-1}\Omega_1$. We then have
\[  \Lambda \, \be_1-JI \Lambda \, \be_2 + J\be_3 = J  \bM+I \,\bN, \ \ {\rm and} \ \ 
\con \Lambda \, \be_1+ JI\, \con\Lambda \, \be_2  - J\be_3 = -J  \bM+I \,\bN .\]  
Solving these equations for $\bM$,  gives
  \beq  \bM =  \frac{\Lambda -\con\Lambda}{2}J\,\be_1 -\frac{\Lambda +\con\Lambda}{2}\, I\,\be_2 +  \be_3, \quad  \be_3\circ  \bM =1, \quad 
  {\rm and} \quad \bM^2 =1-\Lambda \con \Lambda . \label{Msquared} \eeq
  %and used to further simplify 
    %(\ref{decomposeSE}). 
    
  Writing $\bM =\bx + \be_3 = \sum_{k=1}^3 \alpha_k\be_k $, for $\alpha_k \in \G_3^{0+3}$, where 
  $\bx=\alpha_1\be_1+\alpha_2\be_2$ and $\alpha_3=1$, the $\Omega$-matrix
   (\ref{analogpauli}) of $\bM$ takes the form
  \[ [\bM ]_\Omega = \pmatrix{J& \alpha_1 - i \alpha_2 \cr \alpha_1 + i \alpha_2 &- J}_\Omega . \]
  If for the complex idempotent $P$, given in (\ref{complexidempotent}), we make the substitution
  $\bM= \bm$ and $\bN = \bn$ for the vectors $\bm, \bn \in \G_3$, then the idempotent
  $P =\frac{1}{2}(1+ J \bm +I \bn)$. Letting $i\to I$, so that $J \to 1$, the idempotent $P$ becomes
  $s = \frac{1}{2}(1+\bm + I \bn)$, which is the idempotent that was studied in the case of the representation of
  the Pauli spinors on the Riemann sphere, \cite[(23)]{S2014}.   
 
  The complex unit vector $\hat \ba$, defined in (\ref{boostsphere}), can be directly expressed in terms of $\Lambda$. Using 
  (\ref{Msquared}), we find that
\[  \hat \ba =\hat \bM \be_3 \hat \bM =(-\be_3 \hat \bM+2 \be_3\circ \hat \bM)\hat \bM = -\be_3+2 (\be_3 \circ \hat \bM)\hat \bM \]
\[= \frac{2}{ 1 - \Lambda \con \Lambda}\, \bM - \be_3 , \]
  which in turn gives the projective relation
    \beq \bM=\bx+ \be_3  = \frac{1-\Lambda \con \Lambda }{2}(\hat \ba + \be_3)\quad \iff \quad \bM = \frac{2}{\hat \ba+\be_3}\label{projhyper} \eeq
showing that the projection $\bx$ of $\bM$ onto the complex hyperplane defined by $\be_1, \be_2$, is on the
complex ray extending from the south pole $-\be_3$. This is equivalent to saying that $\bM$ is a complex multiple of
the complex vector $\hat \ba + \be_3$.
      
   Using (\ref{decomposeSE}) and (\ref{Msquared}), we find that
  \[ S_E^2=\Omega_0^2 T_E^+ =\Omega_0^2\, J \, \bM\, E_3^+\]
  or 
 \[ S_E = J \Omega_0 \, \bM \, E_3^+=J \Omega_0 \sqrt{\bM^2}\hat \bM E_3^+
         =J \frac{\Omega_0}{\sqrt{\Omega_0 \con \Omega_0}}\sqrt{\Omega_0 \con \Omega_0-\Omega_1\con \Omega_1}\,\hat \bM\, E_3^+,\]
  so that    
 \beq S_E  =J  e^{\large \omega}\hat \bM E_3^+= \hat \bM \be_3 e^{\omega_1+\omega_2 \be_3}E_3^+ 
     =  e^{\omega_1 +\omega_2 \hat \ba} \hat \ba\,\hat\bM E_3^+=Je^{\large \omega}\hat A_3^+\hat \bM, \label{finalSE} \eeq
  for $\large\omega=\omega_1+\omega_2 J$ where $\omega_k = \phi_k+\theta_k I \in \G_3^{0+3}$ for $k=1,2$, 
  and 
  \[ \hat \ba:=\hat \bM \be_3 \hat \bM, \quad A_3^+ :=\hat \bM E_3^+ \hat \bM. \]
  The $\phi_k, \theta_k\in \R$ are defined in such a way that $ e^{\omega_1}
  =\sqrt{\Omega_0 
  \con \Omega_0-\Omega_1\con \Omega_1}=\sqrt{\det[\Omega]}$, and 
  $e^{J\omega_2}:= \frac{\Omega_0}{\sqrt{\Omega_0 \con \Omega_0}}$. 
  Note that $\hat \bM$ can be further decomposed
  using (\ref{boostsphereM}).
  
  There are two additional canonical forms, derived from (\ref{finalSE}), that are interesting. We have
  \beq S_E=e^{\omega_1}e^{I\, \hat \bb_e z_e} e^{\omega_2 \be_3}E_3^+ = e^{\omega_1 }e^{\omega_2 \hat \ba} e^{I\, \hat \bb_a z_a}  E_3^+,
    \label{finalSEnew} \eeq
 where 
 \[ \hat \bM \be_3 = \hat \bM \circ \be_3 + \hat \bM \otimes \be_3 = \cos z_e + I\, \hat \bb_e \sin z_e = e^{I\, \hat \bb_e z_e} \]
 for $z_e \in \G_{1,3}^{0+4}$, $\hat \bb_e \in \G_{1,3}^2$, and $ \hat \bM \otimes \be_3:= \frac{1}{2} (\hat \bM \be_3 -   \be_3\hat \bM)$ is
 the {\it anti-symmetric product}.   Similarly, 
  \[  \hat \ba \hat \bM  =  \hat \ba\circ \hat \bM + \hat \ba\otimes \hat \bM  = \cos z_a + I\, \hat \bb_a \sin z_a = e^{I\, \hat \bb_a z_a} \]
 for $z_a \in \G_{1,3}^{0+4}$ and $\hat \bb_a \in \G_{1,3}^2$.  Note that $\hat \bb_e$ and $\be_3$ anti-commute, as do
 $\hat \bb_a$ and $\hat \ba$.
 % In \cite[(2.13)]{H85}, Hestenes expresses the spinor $\psi$ for the Dirac equation in the form
 %$\psi = (\rho e^{i \beta})^{\frac{1}{2}}R$. Up to the idempotent $E_3^+$, this is equivalent to our decomposition (\ref{finalSEnew}), with
 %$ (\rho e^{i \beta})^{\frac{1}{2}}=e^{\omega_1}$, and $R=e^{I\, \hat \bb_e z_e} e^{\omega_2 \be_3}$. 
 In our decomposition,
 it is the unit bivector $\hat \bM$ that defines the Lorentz transformation (\ref{boostsphere}) associated with a Dirac spinor.

   Recalling (\ref{twistorrep}), (\ref{decomposeSE}), and (\ref{finalSE}), we define a Pauli E-spinor by
  \beq  |\Omega \rangle_E:= \sqrt{2} (\Omega_0+\Omega_1\be_1)E_3^+=\sqrt{2}\Omega_0 T_E^+
  =\sqrt 2 J  e^{\large \omega}\hat \bM E_3^+ , \label{Espinor} \eeq
  Given the E-spinor $| \Phi \rangle_E =\sqrt 2 J  e^{\large \omega^\prime}\hat \bM^\prime E_3^+ $, its conjugate is specified by
  \[ \langle \Phi |_E=\widetilde{\con{| \Phi \rangle}}_E=\sqrt 2 J e^{\large{ \con\omega}^\prime}  E_3^+ \hat \bM^\prime, \]
  which we use to calculate
 \beq  \langle \Phi |_E |\Omega \rangle_E = 2e^{\large{\con \omega}^\prime+\large \omega}E_3^+\hat \bM^\prime \hat \bM E_3^+ 
           = 2e^{\large{\con \omega}^\prime+\large \omega}E_3^+\Big(\hat  \bM^\prime \circ \hat \bM +
            J  (\hat\bM^\prime \otimes \hat \bM) \circ \be_3\Big),  \label{specialcondition} \eeq  
 which can be expressed in the alternative form, 
 \beq   \langle \Phi |_E |\Omega \rangle_E = 2e^{\large{\con \omega}^\prime+\large \omega}\hat \bM^\prime \hat B_3^+ \hat A_3^+ \hat \bM 
       \quad \iff \quad  \langle \Omega |_E |\Phi \rangle_E 
       = 2e^{ \large{\con \omega}+\large \omega^\prime}\hat \bM \hat A_3^+ \hat B_3^+ \hat \bM^\prime,    \label{Etwistorinnerproduct} \eeq         
  and
  \beq  |\Omega \rangle_E  \langle \Omega |_E =2 e^{2\omega_1}\hat \bM E_3^+ \hat \bM = 2 e^{2 \omega_1}\hat A_3^+.
   \label{idempotentA} \eeq
            
    The result (\ref{Etwistorinnerproduct}) is in agreement with (\ref{detomega}) when $\Phi = \Omega$, and identical in terms
  of the complex components $\phi_k$ and $\varphi_k$, for which
  \[  \langle \Phi |_E|\Omega \rangle_E   =2 E_3^+\Big(\con \phi_1 \varphi_1  + \con \phi_2\varphi_2
   -\con \phi_3\varphi_3 -\con \phi_4 \varphi_4
   +  2 im(\con\phi_1\varphi_3+\con \phi_2\varphi_4)I\Big) ,\]
   so the Dirac inner product (\ref{sesquip}) can also be expressed by
   \beq \langle \Phi |\Omega \rangle = \Big \langle \langle \Phi|_E|\Omega\rangle_E\Big\rangle_\C
   =\frac{1}{4}  \Big( \langle \Phi|_E|\Omega\rangle_E+ \langle \Phi|_E|\Omega\rangle_E^-
+ \langle \Phi|_E|\Omega\rangle_E^\dagger+ \langle \Phi|_E|\Omega\rangle_E^*\Big)    , \label{sesquipE}  \eeq
%   where $\langle * \rangle_\S$ takes the $4$-scalar part of the geometric number denoted by $*$. 
Of course, the $\langle \, . \, \rangle^-,\ \langle \, . \, \rangle^\dagger $ and $\langle \, . \, \rangle^*$-conjugation operators all
   depend upon the decomposition of $I=\be_{123}$ in the rest-frame $\{\be_1,\be_2,\be_3\} \in \G_{1,3}^+$. 
%It is highly significant to be able to express the Dirac sesquilinear inner product in terms of Pauli $E$-spinors using the $4$-scalar part, as we will soon discover.  
The importance of this result is that we are able to directly define the Dirac inner product in terms of conjugations of the
Pauli E-inner product given in (\ref{Etwistorinnerproduct}). 
It is also interesting to compare this result with (\ref{sesquipre}).
   
    By a {\it gauge transformation} of an E-spinor $| \Omega \rangle_E $, we
   mean the spinor $e^\alpha | \Omega \rangle_E$, where $\alpha = I \theta + J \phi$ for $\theta, \phi \in \R$.
   Now note that 
   \[ \det [e^\alpha \Omega] =\det \pmatrix{e^\alpha \Omega_0 & e^{\con \alpha}\,\con \Omega_1 \cr e^\alpha  \Omega_1
    & e^{\con \alpha}\, \con \Omega_0} = e^{2I \theta} \det[\Omega] = e^{2I \theta} \Omega_0\con \Omega_0 \bM^2 ,\]
    since $\det[\Omega] = \Omega_0\con \Omega_0 \bM^2$.
  We say that  $|\Omega^\prime\rangle_E =e^\alpha | \Omega \rangle_E$  is 
   {\it gauge normalized} if 
   \beq \det[\Omega^\prime ] \in \R, \ \ e^{i \phi} := \frac{\Omega_0^\prime}{\sqrt{\Omega_0 ^\prime\con \Omega_0^\prime}}\in \C.  \label{gaugenormal} \eeq 
   
  The definition of an E-spinor is closely related to the Dirac g-spinor definition (\ref{twistspin}),
 \beq |\Omega\rangle = \sqrt{2}\,|\Omega \rangle_E \,\gamma_0^+ =2  J  e^{\large \omega}\hat \bM E_3^+\gamma_0^+.
   \label{twistEspinor} \eeq 
 Given the g-spinor $|\Phi\rangle=2  J  e^{\large \omega^\prime}\hat \bM^\prime E_3^+\gamma_0^+ $, its conjugate is    
 \[ \langle \Phi|=\sqrt 2 \gamma_0^+\langle \Phi|_E = 2\gamma_0^+ Je^{{\large \con\omega^\prime}}E_3^+\hat \bM^\prime   . \]
  Using (\ref{Etwistorinnerproduct}) and (\ref{sesquipE}), we calculate 
 \beq \langle \Phi||\Omega \rangle = 2\gamma_0^+ \langle \Phi|_E | \Omega\rangle_E \gamma_0^+
 =4\gamma_0^+e^{\large\con\omega^\prime+\large \omega}E_3^+  \hat \bM^\prime \hat \bM E_3^+\gamma_0^+.\label{usefulstep} \eeq
The Dirac inner product, expressed in (\ref{sesquipE}), shows that
 \beq \langle \Phi|\Omega \rangle = \big\langle \langle \Phi||\Omega \rangle\big\rangle_\C=
  2\big\langle \gamma_0^+ \langle \Phi|_E | \Omega\rangle_E \gamma_0^+ \big\rangle_\C . \label{sesquipinE} \eeq
 %  The powerful decomposition (\ref{finalSE}), and the close relationship between (\ref{Espinor}) and (\ref{twistEspinor}),
 %shows that the study of Dirac g-spinors, can be expressed entirely in terms of Pauli E-spinors. 
 
 Using (\ref{detomega}), (\ref{finalSE}), and (\ref{Etwistorinnerproduct})
  for $|\Omega\rangle_E =\sqrt 2 J  e^{\large \omega}\hat \bM E_3^+ $, 
  \beq  \langle \Omega |_E |\Omega\rangle_E=2e^{2\omega_1}E_3^+\hat \bM \hat \bM E_3^+ 
   =2 \det[\Omega] E_3^+,   \label{innerproductE} \eeq
 so 
 \[  \langle \Omega |\Omega\rangle_E :=  \rho_1^2 e^{2I\theta_1} = \det[\Omega], \]
 where $\rho_1 := e^{\phi_1}$. When $|\Omega \rangle_E$ is gauge normalized
 to 
 \[ |\Omega^\prime\rangle_E:=e^{-I \theta_1 - J\phi_2} |\Omega \rangle_E, \]
 and similarly, $|\Phi \rangle_E$ is gauge normalized to $|\Phi^\prime \rangle_E$ , then
 the inner products (\ref{sesquipinE}) and (\ref{innerproductE})  become more simply related. For
 gauge normalized E-spinors $|\Omega\rangle_E$ and $|\Phi \rangle_E$, the relation (\ref{usefulstep}) simplifies to
 \beq   \langle \Phi||\Omega \rangle = 2\gamma_0^+ \langle \Phi|_E | \Omega\rangle_E \gamma_0^+
 =4e^{\large\con\omega^\prime+\large \omega}\gamma_0^+E_3^+  \hat \bM^\prime \hat \bM E_3^+\gamma_0^+ .\label{normalizedinner} \eeq
 A geometric Dirac spinor $|\Omega\rangle$ represents the {\it physical state} of a spin $\frac{1}{2}$-particle if 
 $\det[\Omega]=1$.
 
  Before our next calculation, for $\hat A_+ = \frac{1}{2}(1+J \hat \ba)$ and $\hat B = \frac{1}{2}(1+J\,\hat \bb)$
  for $\hat \ba, \hat \bb \in \G_{1,3}^{2}$, we find that
\[ \hat A_+ \hat B_+ \hat A_+= \frac{1}{4} \hat A_+(1+\hat B_E)(1+\hat A_E) =
    \frac{1}{4} \hat A_+(1+\hat B_E +\hat A_E + \hat B_E \hat A_E )  \]
   \beq  =  \frac{1}{4} \hat A_+(1+\hat B_E +\hat A_E - \hat A_E \hat B_E +2 \hat A_E \circ B_E  ) = 
     \frac{1}{2} \hat A_+ (1+ \hat \ba \circ \hat \bb) .\label{tripleidempotents}   \eeq 
  Using this result, (\ref{Etwistorinnerproduct}) and (\ref{sesquipE}), we calculate
    \[\langle \Omega |_E|\Phi \rangle_E  \langle \Phi |_E|\Omega \rangle_E 
  = 4e^{2 (\omega_1+\omega_1^\prime)} \hat \bM_\Omega \hat A_\Omega^+ \hat A_\Phi^+ \hat A_\Omega^+ 
 \hat \bM_\Omega \] 
 \beq= 2e^{2 (\omega_1+\omega_1^\prime)} 
  (1+\hat \ba_\Phi \circ \hat \ba_\Omega)E_3^+ = 
   2 \det[\Omega] \det[\Phi] (1+\hat \ba_\Phi \circ \hat \ba_\Omega)E_3^+ .  \label{probabilityAB} \eeq  
   
Using (\ref{sesquipE}), we would now like to calculate $ \langle \Omega | \Phi \rangle\langle \Phi | \Omega \rangle $ for gauge normalized physical states $|\Omega\rangle_E$ and $|\Phi\rangle_E$. Here, we consider a special case
where the formula (\ref{sesquipE}) simplifies to
 \beq \langle \Phi|\Omega \rangle =\frac{1}{2}  \Big( \langle \Phi|_E|\Omega\rangle_E+ 
 \langle \Phi|_E|\Omega\rangle_E^\dagger\Big)    . \label{sesquipinEsimple} \eeq
Referring to (\ref{specialcondition}), this will occur when 
 \beq \hat \bM_\Phi\circ \hat \bM_\Omega =( \hat \bM_\Phi \circ \hat \bM_\Omega )^\dagger \quad {\rm and} \quad
         (\hat\bM_\Phi \otimes \hat \bM_\Omega) \circ \be_3 =-\Big(\big(\hat\bM_\Phi \otimes \hat \bM_\Omega) \circ \be_3\Big)^\dagger  . \label{specialconditionused} \eeq  

Indeed, if $|\Omega\rangle_E$ and $|\Phi\rangle_E$ are gauge normalized spinor states, 
   then using (\ref{sesquipE}) and (\ref{probabilityAB}), we have
   \[ \langle \Omega | \Phi \rangle\langle \Phi | \Omega \rangle 
   =\frac{1}{4}  \Big( \langle \Omega |_E|\Phi \rangle_E + \langle \Omega |_E|\Phi \rangle_E^\dagger\Big)\Big(
     \langle \Phi |_E|\Omega \rangle_E + \langle \Phi |_E|\Omega \rangle_E^\dagger\Big)  \]
     \beq =  \frac{1}{2} \det[\Omega] \det[\Phi] \big(1+\hat \ba_\Phi \circ \hat \ba_\Omega\big). \label{normalizedprob}   \eeq
     If $\det[\Omega] =\det[\Phi]=1$, then the probability of finding the spin $\frac{1}{2}$-particle in the state $|\Phi\rangle$, 
     having prepared the particle in the 
     state $|\Omega \rangle$, is given by $\langle \Omega | \Phi \rangle\langle \Phi | \Omega \rangle $.    
  There is a particularly simple formula for evaluating (\ref{normalizedprob}), directly in terms of $ \bM_\Omega$ and
$\bM_\Phi$, which follows from (\ref{projhyper}). We find that
\beq   \langle \Omega | \Phi \rangle\langle \Phi | \Omega \rangle  =\frac{1}{2} \big(1+\hat \ba_\Phi \circ \hat \ba_\Omega\big)= 1 -  \frac{(\bM_\Omega - \bM_\Phi)^2}{{\bM_\Omega^2 \bM_\Phi^2}} . 
                  \label{normalizedprobnew}   \eeq
 
   Let us consider important examples of gauge normalized states for which (\ref{normalizedprobnew}) applies. 
Let 
\beq  \bM_\Omega =e^{\phi_x \be_3}\bx+\be_3 = \bx \cosh\phi_x + \be_3+ I \be_3 \times \bx \sinh \phi_x , \label{notsurprise} \eeq
for $\bx=x_1 \be_1+x_2 \be_2$ and $\phi_x \in \R$. This $\bM_\Omega$ is defined by
$\Lambda=\Omega_0^{-1} \Omega_1$, for
\[ \Omega_0 = \varphi_1 + J \varphi_3 = 1, \  \ {\rm and} \  \  \Omega_1= \varphi_4+ J \varphi_2= J e^{-J \phi_x}(x_1+x_2i), \]
or alternatively by
\[ \Omega_0 = \varphi_1 + J \varphi_3 = e^{\phi_x}, \  \   {\rm and} \  \  \Omega_1= \varphi_4+ J \varphi_2= J (x_1+x_2 i) .\]

%The usual {\it up} state, with $\Omega_0=1, \Omega_1=0$ See Figure \ref{example2}.

For $\bM_{\Omega}$, we calculate $\bM_{\Omega}^2=\bx^2+1\ge 1$, and
\beq \hat \bM_{\Omega}=  \frac{\bx \cosh \phi_x+\be_3}{\sqrt{\bx^2+1}}+ 
  \frac{I(\be_3\times \bx)\sinh \phi_x}{\sqrt{\bx^2+1}} = \bm_1+I \bm_2 , \label{m1m2form} \eeq
which defines the velocity 
\[ \tanh\frac{\omega_x}{2} ={\frac{v}{c}}:= \sqrt{\frac{(\bm_2)^2}{(\bm_1)^2}}= \sqrt{\frac{\bx^2 \sinh^2 \phi_x}{\bx^2 \cosh^2 \phi_x+1}}  \]
in the direction
\[  \hat \bm_1 \times \hat \bm_2 = \frac{-\hat\bx + \be_3|\bx|\cosh \phi_x }
{\sqrt{\bx^2 \cosh^2 \phi_x+1}} . \]
Given $\bM_{\Omega}$, we can also find the element 
\[   \bM_{\Omega}^\perp := \frac{1}{\bM_{\Omega} \otimes \be_3}\bM_{\Omega}
 =-\frac{1}{e^{\phi_x \be_3}\be_3} \bM_{\Omega}
 \]
\beq =\be_3 \bx^{-1} e^{-\phi_x \be_3}\big(e^{\phi_x \be_3}\bx +\be_3\big)= -e^{\phi_x \be_3} \frac{\bx }{\bx^2}+ \be_3, \label{perpMa} \eeq
for which $\hat \ba_x^\perp = - \hat \ba_x$. Notice that $\bM_{\Omega}^\perp$ and $\bM_{\Omega}$ are in the same inertial system defined by $\phi_x$.

For $\bM_{\Omega}$, and a second state $\bM_{\Phi} =e^{\phi_y \be_3}\by+\be_3$, we now calculate
\[  \frac{1}{2}(1+\hat \ba_x \circ \hat \ba_y)= 1-\frac{(\bM_{\Omega}-\bM_{\Phi})^2}{\bM_{\Omega}^2 \bM_{\Phi}^2} \]
\beq =
    1- \frac{\bx^2+\by^2-2\Big( \cosh(\phi_x-\phi_y)\bx\cdot \by + \sinh(\phi_x-\phi_y)\be_3\w\bx\w \by\Big)}{\Big(1+\bx^2\Big)\Big(1+\by^2\Big)} .\label{probadotbcomplex} \eeq

\begin{figure}
\begin{center}
\no\includegraphics[scale=0.75]{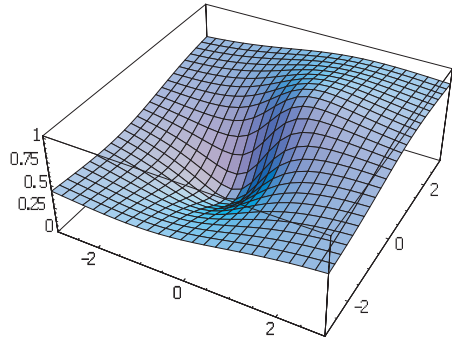}
\caption{Shown is the discrete probability distribution for observing a particle in a state $\hat \ba_x$, given that
the particle was prepared in the state $\hat \ba_y$ immediately preceding. For this figure, $\by=(1 ,1)$,
$\phi_x=\phi_y=0$, $-3\le x_1 \le 3$ and $-3\le x_2 \le 3$.}
\label{example1}
\end{center}
\end{figure}

When $\phi_x=\phi_y$, so that $\bM_{\Omega}$ and $\bM_{\Phi}$ are in the same inertial system, the formula (\ref{probadotbcomplex}) reduces to
\beq  \frac{1}{2}(1+\hat \ba_x \circ \hat \ba_y) = \frac{\bx^2 \by^2+2 \bx \cdot \by+1}{(\bx^2+1)(\by^2+1)}, 
\label{probadotbreal} \eeq
which is the discrete Pauli probability of measuring the state $\bM_\Omega$, given the state $\bM_\Phi$, and it is independent of the inertial system in which it
is calculated \cite{S2014}. Note that the inertial system defined by $\phi_x$ is different than the inertial system of $\hat \ba_x$ and $\hat \ba_y$ defined by $\omega_x$. See Figure \ref{example1}. 

\begin{figure}
\begin{center}
\no\includegraphics[scale=0.75]{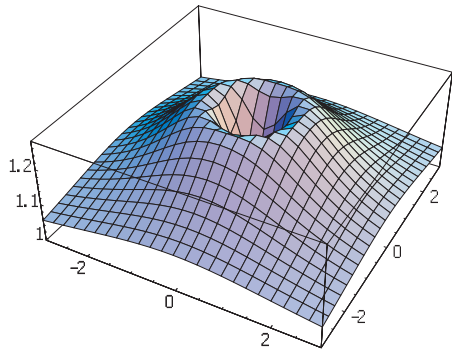}
\caption{Shown is the discrete distribution for observing a particle in a state $\hat \ba_x$, given that
the particle was prepared in the state $\hat \ba_y$ immediately preceding. For this figure, $\bx=\by$,
$\phi_x=1, \ \phi_y=2$, $-3\le x_1 \le 3$ and $-3\le x_2 \le 3$.
The values are all greater than or equal to one and therefore do not represent a probability.} 
\label{example2}
\end{center}
\end{figure}

Also, when $\by= \rho \bx$, the pseudoscalar term will vanish. For example,
for $\bx=\by$, and $\phi_x=1, \ \phi_y=2$, the formula (\ref{probadotbcomplex}) gives
 \[ \frac{1}{2}(1+\hat \ba_x \circ \hat \ba_y) =  1+2 \bx^2 \frac{\big(\cosh(1) -1\big)}{(1+\bx^2)^2}\ge1,  \]
which cannot represent a probability distribution, but gives the volcano shown in Figure \ref{example2}. 
In the general
case for $\bM_{\Omega}$ and $\bM_{\Phi}$, the term $\bM_{\Omega}\circ \bM_{\Phi}$ will contain the
 pseudoscalar element $  \be_3\w \bx \w \by$,
which vanishes in the special cases considered above. 
 
There is one other spin state that we will consider. Let
\[  \bM = \bx(1+\be_3)+\be_3=x \be_1+y \be_2+ \be_3+I(y \be_1-x \be_2)= \hat \bM . \]
This state is realized for $\varphi_1=1, \varphi_3=0$ and $\varphi_2=\varphi_4=x+iy$, so that
$\Lambda=(1+J)(x+iy)$. We then calculate
     \[  \hat \ba = \hat \bM \be_3 \hat \bM = -\be_3 + 2(\be_3 \circ \hat \bM ) \hat \bM 
   =2 \bx(1+\be_3)+\be_3. \]
This state is a limiting state of (\ref{notsurprise}), when $\phi_x \to -\infty$, and cannot
represent the state of a spin $\frac{1}{2}$ particle.

% We give one more example of a discrete probability distribution. In this case, we measure
%a particle prepared in the spin-up state defined by $\bM_\Phi =x_1\be_1+ \be_3$, in the spin state defined by
%$\bM_\Omega= \bx+ \be_3$. We find that
%\[ \langle \Omega | \Phi \rangle\langle \Phi | \Omega \rangle  =\frac{1}{2} \big(1+\hat \ba_\Phi \circ \hat \ba_\Omega\big)= 1 -  \frac{(\bM_\Omega - \bM_\Phi)^2}{\bM_\Omega^2 \bM_\Phi^2}=1-\frac{x_2^2}{(x_1^2+1)\bx^2}. \]
%The discrete distribution is shown in Figure \ref{example4}.

 \section{Fierz Identities}
 
 The so-called Fierz identities are quadratic relations between the {\it physical observables} of a Dirac spinor.
 The identities are most easily calculated in terms of the spinor operator $\psi$ of a Dirac spinor. Whereas spinors are usually classified
 using irreducible representations of the Lorentz group $SO_{1,3}^+$, Pertti Lounesto has developed a classification scheme
 based upon the Fierz identities \cite[P.152,162]{LP97}. A Dirac spinor, which describes an electron, the subject of this paper, is characterized 
 by the property that $\det[\psi]_\Omega \ne 0$. Other types of spinors, such as Majorana and Weyl spinors, and even Lounesto's {\it boomerang spinor},
 can all be classifed by bilinear covariants of their spinor operators $\psi \in \G_3\widetilde = \G_{1,3}^+$.

  %Recall that $\G_3 \widetilde= \G_{1,3}^+$, the even sub-algebra of $\G_{1,3}$, from which it follows that
  %\[\G_3(\C)\widetilde=\G_{1,3}^+(\C). \]
    For $g\in \G_3$, let $g^-$ and $g^\dagger$,
and $g^*$ denote the conjugations of {\it inversion}, {\it reversion}, and their composition $g^*= ({g^-})^\dagger$,
respectively. Thus for $g=\alpha+\bx+ I \, \by  $, where $\alpha \in \G^{0+3}$,
\beq g^-= \alpha^\dagger - \bx + I\, \by,\quad g^\dagger =  \alpha^\dagger +\bx - I\, \by, \quad {\rm  and} \quad g^* =\alpha - \bx- I\, \by. 
\label{pauliconjugates} \eeq
Any element $\omega \in \G_{1,3}$ can be written $\omega = g_1 + g_2 \gamma_0$ for some $g_1,g_2 \in \G_3$.
For $\omega \in \G_{1,3}$, as before, let $\widetilde \omega$ denote the conjugation of reverse 
in $\G_{1,3}$. Since $g^\dagger =\gamma_0g^*\gamma_0$,
\[ \widetilde \omega = \widetilde g_1 + \gamma_0 \widetilde g_2 = g_1^* + \gamma_0 g_2^*=g_1^*+g_2^\dagger \gamma_0 .\]

Each unit timelike vector $\gamma_0\in \G_{1,3}$ determines a unique Pauli sub-algebra $\G_3 \widetilde =\G_{1,3}^+$, and
its corresponding {\it rest-frame} $\{\be_1,\be_2,\be_3\}$, satisfying $\be_k \gamma_0 =-\gamma_0 \be_k$ for
$k=1,2,3$. The relative conjugations of the Pauli algebra $\G_3$ can be defined directly in terms of the the conjugations
of the larger algebra $\G_{1,3}$. For $g\in \G_3$,
\[ g^-:=\gamma_0 g \gamma_0,\quad g^\dagger:=\gamma_0 \widetilde g \gamma_0, \quad g^*:= (g^-)^\dagger = \widetilde g. \]
Using (\ref{pauliconjugates}), we calculate
\[  g g^\dagger = \alpha  \alpha^\dagger + \bx^2+\by^2+(\alpha +  \alpha^\dagger)\bx-(\alpha-  \alpha^\dagger)I\, \by =r +s \hat \ba\in \G^{0+1},\]
\[  gg^* = (\alpha + \bx+ I\, \by)(\alpha -\bx - I\, \by) = \alpha^2-(\bx - I\, \by)^2 =\det [g]_\Omega \in \G_3^{0+3}, \]
or $g g^* = R_1 + I\, R_2$ for $R_1,R_2 \in \R$.
    
   Define
   \[ \bJ := g \gamma_0 g^* = g \gamma_0 g^* \gamma_0 \gamma_0 = g g^\dagger \gamma_0= \theta \gamma_0 + \phi \hat a , \]
 where $\hat a^2 =-1$ and $\gamma_0 \cdot \hat a = 0$,
   \[ \bS := g \gamma_{12}g^*=-I g \be_3 g^*, \quad {\rm and} \quad \bK:=g \gamma_3 g^*=g \be_3 g^\dagger \gamma_0 .  \]   
  We then find
  \[ \bJ^2 =g \gamma_0 g^* g \gamma_0 g^* =  \big(R_1+ I R_2\big)\big( R_1- I R_2 \big) =
                 R_1^2+  R_2^2 =r^2-s^2 \ge 0 ,\]
   $\bS$ is a bivector in $\G_{1,3}$ since $\widetilde\bS=\bS^*=-\bS$, and
   \[ \bS^2=\big(-Ig\be_3g^*\big)\big(-Ig\be_3g^*\big)=-g \be_3 R \be_3 g^* =- R^2 \]
   where $R=R_1+I R_2$. Similarly, $\widetilde \bK =\bK$ and the {\it inverse} of $\bK\in \G_{1,3}$ is $-\bK$,
   from which it follows that $\bK\in \G_{1,3}^1$, and 
   \[ \bK^2 = g \gamma_3 g^*g \gamma_3 g^*=g\gamma_3 R \gamma_3 g^* =- R  R^\dagger = - \bJ^2\le 0. \]  
   
   We also find that
    \[ \bK \bJ = g \gamma_3 g^* g\gamma_0 g^* = g \gamma_3 R \gamma_0 g^*=g \be_3 g^*  R^\dagger = I \bS   R^\dagger
       = \bK \w \bJ, \]
 and consequently, $\bK \cdot \bJ = 0$. Note also that
   \[ \bJ \bS = -I g \gamma_0 g^*g\be_3 g^* = -I g \gamma_0 R \be_3 g^*= I  R^\dagger \bK ,\]
 and
  \[ \bJ \bS \bK=I  R^\dagger\bK^2=-I |R|^2  R^\dagger. \]                   
                 
Writing $g=(x_0+\bx)+I(y_0+\by)$, $g^\dagger =(x_0+\bx)-I(y_0+\by)$, and
$g^*= (x_0-\bx)-I(y_0-\by)$, calculate
 \[ g g^\dagger =(x_0^2+y_0^2+\bx^2+\by^2)+2(x_0\bx+y_0\by+\bx \times \by), \]
 \[ g^\dagger g =(x_0^2+y_0^2+\bx^2+\by^2)+2(x_0\bx+y_0\by-\bx \times \by),\]
and
\[ g g^* =g^*g =x_0^2-y_0^2 +\by^2-\bx^2+2I\big( x_0y_0-\bx \cdot \by\big) = R.\]
We see that
\[ \bJ=gg^\dagger \gamma_0=\theta\gamma_0+\phi \hat a =(x_0^2+y_0^2+\bx^2+\by^2)\gamma_0+2(x_0\bx+y_0\by+\bx \times \by)\gamma_0. \]

%\end{document}     
 \section*{Acknowledgements} I thank Professor Melina Gomez Bock  and her students for many lively discussions about quantum mechanics, and the Universidad de Las Americas-Puebla for many years of support. Professor Leonard Susskind's
You Tube Lectures  \cite{S07}, inspired me to look closely at the derivation of the discreet probability distribution of the
Pauli spin states of an electron.

\end{document}